\documentclass[runningheads]{llncs}
\usepackage[T1]{fontenc}
\usepackage{graphicx}
\usepackage{graphicx}
\usepackage{amsmath}
\usepackage{dsfont}
\usepackage{subcaption}
\usepackage{easyReview}
\usepackage{hyperref}
\usepackage{multirow}
\usepackage{booktabs}

\usepackage{color}

\begin{document}
\title{unORANIC: Unsupervised Orthogonalization of Anatomy and Image-Characteristic Features}
\titlerunning{unORANIC}
%
\author{Sebastian Doerrich \and
Francesco Di Salvo \and
Christian Ledig}
\authorrunning{S. Doerrich et al.}
%
\institute{xAILab, University of Bamberg, Germany \\
\email{sebastian.doerrich@uni-bamberg.de}}
\maketitle              
\begin{abstract}
We introduce unORANIC, an unsupervised approach that uses an adapted loss function to drive the orthogonalization of anatomy and image-characteristic features. The method is versatile for diverse modalities and tasks, as it does not require domain knowledge, paired data samples, or labels. During test time unORANIC is applied to potentially corrupted images, orthogonalizing their anatomy and characteristic components, to subsequently reconstruct corruption-free images, showing their domain-invariant anatomy only. This feature orthogonalization further improves generalization and robustness against corruptions. We confirm this qualitatively and quantitatively on 5 distinct datasets by assessing unORANIC's classification accuracy, corruption detection and revision capabilities. Our approach shows promise for enhancing the generalizability and robustness of practical applications in medical image analysis. The source code is available at \href{https://github.com/sdoerrich97/unORANIC}{github.com/sdoerrich97/unORANIC}.

\keywords{Feature Orthogonalization \and Robustness \and Corruption Revision \and Unsupervised learning \and Generalization}
\end{abstract}
\section{Introduction}
\begin{figure}[t]
\centering
\begin{subfigure}{0.48\textwidth}
    \includegraphics[width=\textwidth]{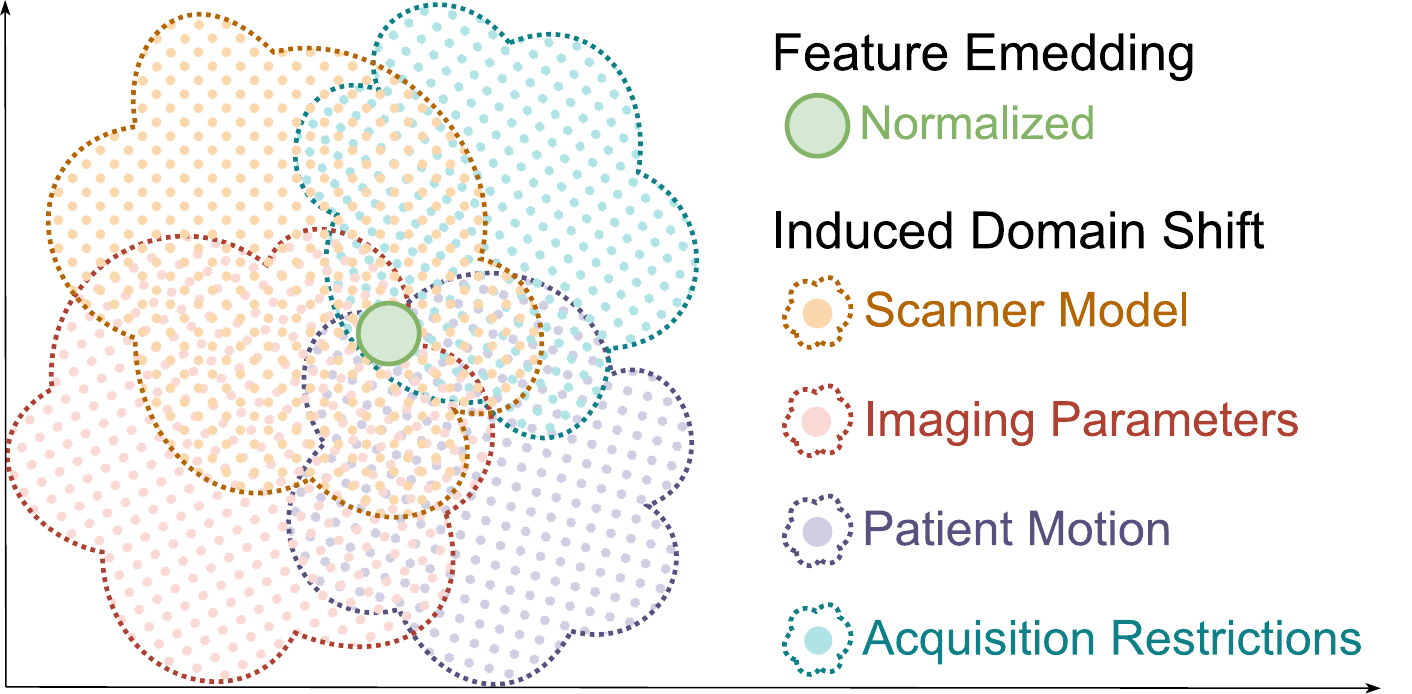}
    \caption{}
    \label{fig:motivation_domainShift}
\end{subfigure}
\begin{subfigure}{0.48\textwidth}
    \includegraphics[width=\textwidth]{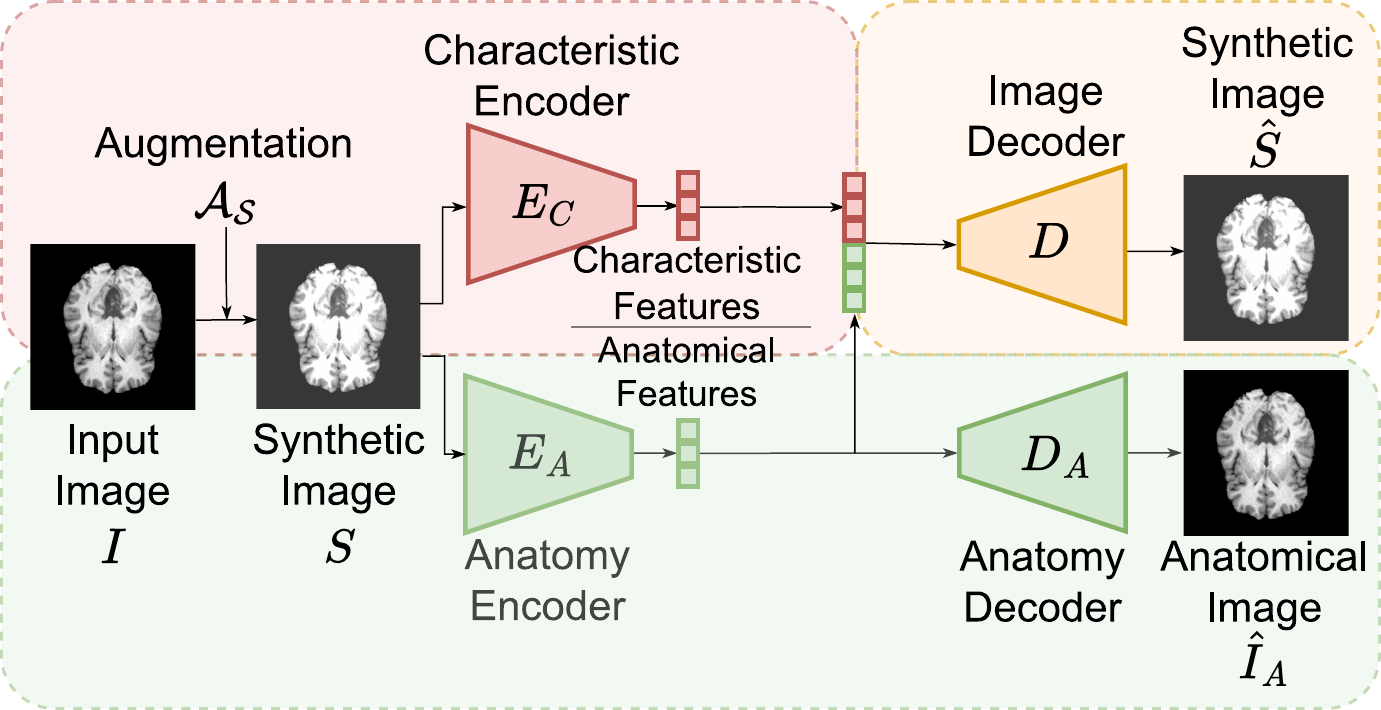}
    \caption{}
    \label{fig:simplified_architecture}
\end{subfigure}
\caption{\textbf{a)} Schematic visualization of a domain disparity caused by variations in the imaging process. \textbf{b)} Visualization of unORANIC's fundamental idea.}
\label{fig:methods}
\end{figure}

In recent years, deep learning algorithms have shown promise in medical image analysis, including segmentation \cite{Ronneberger2015,Rondinella2023}, classification \cite{He2016,MANZARI2023106791}, and anomaly detection \cite{Ngo2019,Jeong2023CVPR}. However, their generalizability across diverse imaging domains remains a challenge \cite{Eche2021} especially for their adoption in clinical practice \cite{Stacke2021} due to domain shifts caused by variations in scanner models \cite{Khan2022}, imaging parameters \cite{Lafarge2017}, corruption artifacts \cite{Priyanka2020}, or patient motion \cite{Oksuz2020}. A schematic representation of this issue is presented in \figurename~\ref{fig:motivation_domainShift}. To address this, two research areas have emerged: domain adaptation (DA) and domain generalization (DG). DA aligns feature distributions between source and target domains \cite{Li2021}, while DG trains models on diverse source domains to learn domain-invariant features \cite{LiDa2018}. Within this framework, some methods aim to disentangle anatomy features from modality factors to improve generalization. Chartsias et al. introduce SDNet, which uses segmentation labels to factorize 2D medical images into spatial anatomical and non-spatial modality factors for robust cardiac image segmentation \cite{Chartsias2019}. Robert et al. present HybridNet that utilizes a two-branch encoder-decoder architecture to learn invariant class-related representations via a supervised training concept \cite{Robert2018}. Dewey et al. propose a deep learning-based harmonization technique for MR images under limited supervision to standardize across scanners and sites \cite{Dewey2020}. In contrast, Zuo et al. propose unsupervised MR image harmonization to address contrast variations in multi-site MR imaging \cite{Zuo2021} by using T1- and T2-weighted image pairs of the same patient. However, all of these approaches are constrained by either requiring a certain type of supervision, precise knowledge of the target domain, or specific inter-/intra-site paired data samples.

To address these limitations, we present unORANIC, an approach to orthogonalize anatomy and image-characteristic features in an unsupervised manner, without requiring domain knowledge, paired data, or labels of any kind. The method enables bias-free anatomical reconstruction and works for a diverse set of modalities. For that scope, we jointly train two encoder-decoder branches. One branch is used to extract true anatomy features and the other to model the characteristic image information discarded by the first branch to reconstruct the input image in an autoencoder objective. A high-level overview of the proposed approach is provided in \figurename~\ref{fig:simplified_architecture}. Experimental results on a diverse dataset demonstrate its feasibility and potential for improving generalization and robustness in medical image analysis.

\section{Methodology}
For the subsequent sections, we consider the input images as bias-free and uncorrupted ($I$). We further define $\mathcal{A_S}$ as a random augmentation that distorts an input image $I$ for the purpose to generate a synthetic, corrupted version $S$ of that image. As presented in \figurename~\ref{fig:simplified_architecture}, such a  synthetic image $S$ is obtained via the augmentation $\mathcal{A_S}$ applied to $I$ and subsequently fed to both the anatomy encoder $E_A$ and the characteristic encoder $E_C$ simultaneously. The resulting embeddings are concatenated $\left( \oplus \right)$ and forwarded to a convolutional decoder $D$ to create the reconstruction $\hat{S}$ with its specific characteristics such as contrast level or brightness. By removing these characteristic features in the encoded embeddings of $E_A$, we can reconstruct a distortion-free version ($\hat{I}_A$) of the original input image $I$. To allow this behavior, the anatomy encoder, $E_A$, is actively enforced to learn acquisition- and corruption-robust representations while the characteristic encoder $E_C$ retains image-specific details.

\subsection{Training}
During training the anatomy encoder is shared across the corrupted image $S$ as well as additional corrupted variants $v_i$ of the same input image $I$ to create feature embeddings for all of them. By applying the consistency loss $\mathcal{L}_{\text{C}}$ given as:
\begin{equation}
    \mathcal{L}_{\text{C}} = \frac{1}{Z} \left( \sum_{\forall\,\{v_i, v_j\}\,\in\,V,~v_i \neq v_j} \left|\left| E^{A}(v_i) - E^{A}(v_j) \right|\right|_2 \right) \quad \text{with } Z = \begin{pmatrix}V \\ 2 \end{pmatrix}
\end{equation}
on the resulting feature maps, the anatomy encoder is forced to learn distortion-invariant features. Here, $V$ is the set of all variants including $S$, $v_i$ and $v_j$ are two distinct variants as well as $E^{A}(v_i)$ and $E^{A}(v_j)$ are the corresponding anatomy feature embeddings. To further guide the anatomy encoder to learn representative anatomy features, the consistency loss is assisted by a combination of the reconstruction loss of the synthetic image $\mathcal{L}_{\text{R}_{S}}$ and the reconstruction loss of the original, distortion-free image $\mathcal{L}_{\text{R}_{I}}$ given as:
\begin{equation}
    \mathcal{L}_{\text{R}_{S}} = \frac{1}{N M} \left|\left| S - \hat{S} \right|\right|_2 = \frac{1}{N M} \left|\left| S - D\left(E_A(S)\,\oplus\,E_C(S) \vphantom{\hat{S}} \right) \right|\right|_2
\end{equation}
\begin{equation}
    \mathcal{L}_{\text{R}_{I}} = \frac{1}{N M} \left|\left| I - \hat{I}_A \right|\right|_2 = \frac{1}{N M} \left|\left| I - D_A(E_A(I)) \vphantom{\hat{I}_D} \right|\right|_2
\end{equation}
with $N$ and $M$ being the image height and width respectively. Using $\mathcal{L}_{\text{R}_{I}}$ to update the encoder $E_A$, as well as the decoder $D_A$, allows the joint optimization of both. This yields more robust results than updating only the decoder with that loss term. The complete loss for the anatomy branch is therefore given as:
\begin{equation}
    \mathcal{L}_{\text{total}} = \lambda_{\text{reconstruction}} \left(\mathcal{L}_{\text{R}_{I}} + \mathcal{L}_{\text{R}_{S}}\right) + \lambda_{\text{consistency}}\,\mathcal{L}_{\text{C}}
\end{equation}
where $\lambda_{\text{reconstruction}}$ and $\lambda_{\text{consistency}}$ are two non-negative numbers that control the trade-off between the reconstruction and consistency loss.
In contrast, the characteristic encoder $E_C$ and decoder $D$ are only optimized via the reconstruction loss of the synthetic image $\mathcal{L}_{\text{R}_{S}}$. 

\subsection{Implementation}
\begin{figure}
\centering
\includegraphics[width=0.64\textwidth]{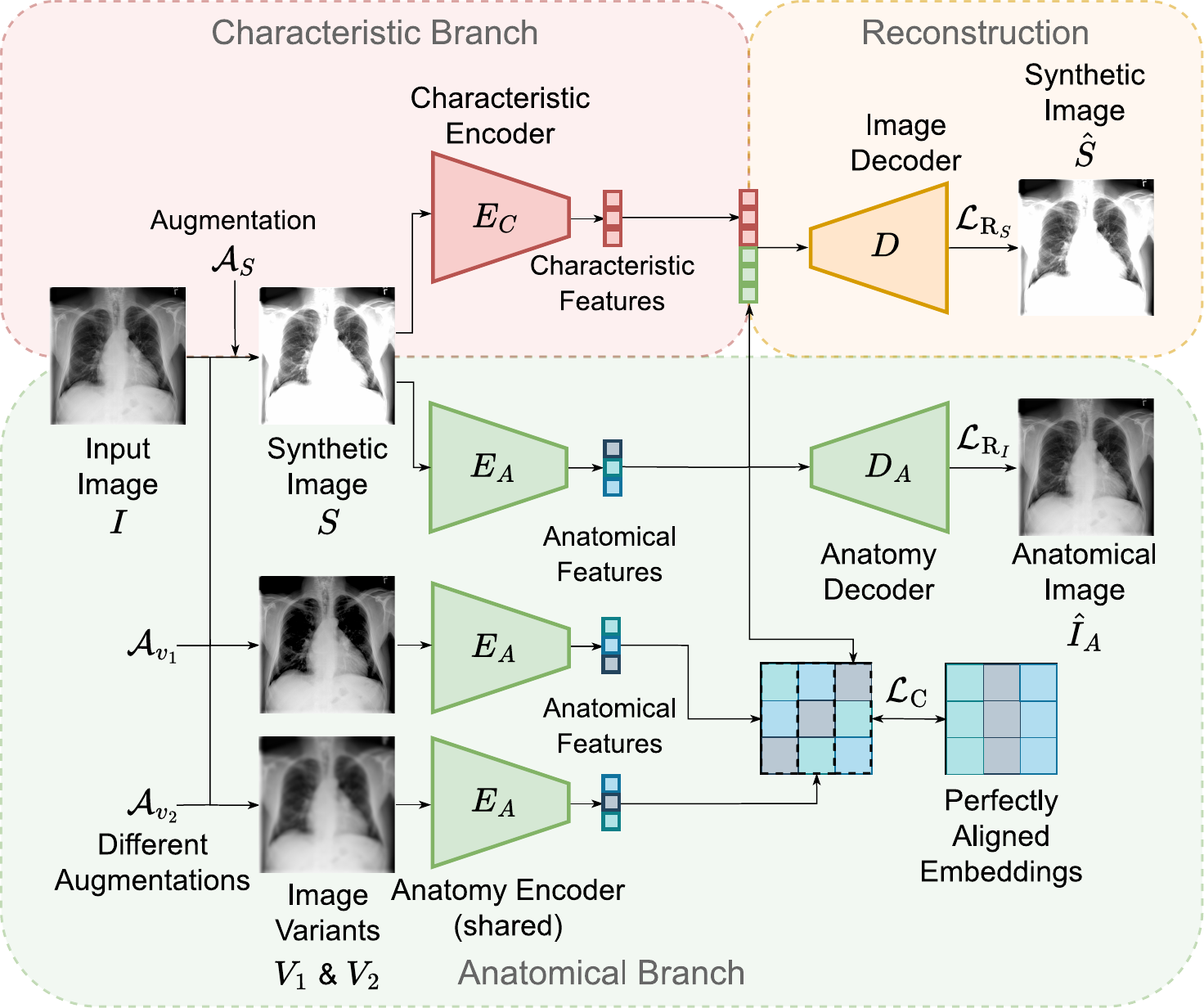}
    \caption{Training pipeline of unORANIC for CT images.}
    \label{fig:architecture}
\end{figure}

The entire training pipeline is depicted in \figurename~\ref{fig:architecture}. Both encoders consist of four identical blocks in total, where each block comprises a residual block followed by a downsampling block. Each residual block itself consists of twice a convolution followed by batch normalization and the Leaky ReLU activation function. The downsampling blocks use a set of a strided convolution, a batch normalization, and a Leaky ReLU activation to half the image dimension while doubling the channel dimension. In contrast, the decoders mirror the encoder architecture by replacing the downsampling blocks with corresponding upsampling blocks for which the convolutions are swapped with transposed convolutional layers. During each iteration, the input image $I$ is distorted using augmentation $\mathcal{A}_S$ to generate $S$ and subsequently passed through the shared anatomy encoder $E_A$ in combination with two different distorted variants $V_1$ and $V_2$. These variants comprise the same anatomical information but different distortions than $S$. The consistency loss $\mathcal{L}_{\text{C}}$ is computed using the encoded feature embeddings, and the reconstruction loss of the original, distortion-free image $\mathcal{L}_{\text{R}_{I}}$ is calculated by passing the feature embedding of $S$ through the anatomy decoder $D_A$ to generate the anatomical reconstruction $\hat{I}_A$. Furthermore, the anatomy and characteristic feature embeddings are concatenated and used by the image decoder $D$ to reconstruct $\hat{S}$ for the calculation of $\mathcal{L}_{\text{R}_{S}}$. $\mathcal{A}_S$, $\mathcal{A}_{v_1}$ and $\mathcal{A}_{v_2}$ are independent of each other and can be chosen randomly from a set of augmentations such as the Albumentations library \cite{Buslaev2020}. The network is trained until convergence using a batch size of $64$ as well as the Adam optimizer with a cyclic learning rate. Experiments were performed on $28 \times 28$ pixel images with a latent dimension of $256$, but the approach is adaptable to images of any size. The number of variants used for the anatomy encoder training is flexible, and in our experiments, three variants were utilized.

\section{Experiments and Results}
We evaluate the model's performance in terms of its reconstruction ability as well as its capability to revise existing corruptions within an image. Further, we assess its applicability for the execution of downstream tasks by using the anatomy feature embedding, as well as its capacity to detect corruptions by using the characteristic feature embedding. Last, we rate the robustness of our model against different severity and types of corruption.

\subsection{Dataset}
To demonstrate the versatility of our method, all experiments were conducted on a diverse selection of datasets of the publicly available MedMNIST v2 benchmark\footnote{Yang, J., et al. \textit{MedMNIST v2-A large-scale lightweight benchmark for 2D and 3D biomedical image classification}. Scientific Data. 2023. License: CC BY 4.0. Zenodo. \url{https://zenodo.org/record/6496656}}\cite{Yang2020,Yang2023}. The MedMNIST v2 benchmark is a comprehensive collection of standardized, 28 x 28 dimensional biomedical datasets, with corresponding classification labels and baseline methods. The selected datasets for our experiments are described in detail in \tablename~\ref{tab:dataset}.

\begin{table}
\caption{Details of the selected set of datasets}
\label{tab:dataset}
    \centering
    \begin{tabular}{l c c c c }
        \toprule
        MedMNIST Dataset & Modality & Task (\# Classes) & \# Train / Val / Test \\
        \midrule
        Blood \cite{Acevedo2020} & Blood Cell Microscope & Multi-Class (8) & $11,959$ / $1,712$ / $3,421$ \\
        Breast \cite{Al-Dhabyani2020} & Breast Ultrasound & Binary-Class (2) & $546$ / $78$ / $156$ \\
        Derma \cite{Tschandl2018,Codella2019} & Dermatoscope & Multi-Class (7) & $7,007$ / $1,003$ / $2,005$ \\
        Pneumonia \cite{Kermany2018} & Chest X-Ray & Binary-Class (2) & $4,708$ / $524$ / $624$ \\
        Retina \cite{Liu2022} & Fundus Camera & Ordinal Regression (5) & $1,080$ / $120$ / $400$ \\
        \bottomrule
    \end{tabular}
\end{table}

\subsection{Reconstruction}
\label{subsec:reconstruction}
\begin{figure}[t]
\centering
\begin{subfigure}{0.32\textwidth}
    \includegraphics[width=\textwidth]{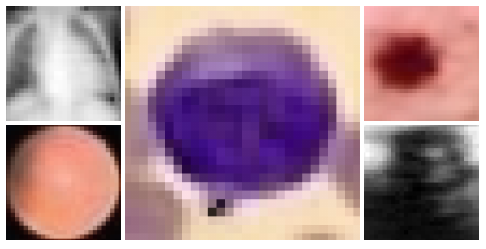}
    \caption{}
    \label{fig:reconstructions_ours}
\end{subfigure}
\hfill
\begin{subfigure}{0.32\textwidth}
    \includegraphics[width=\textwidth]{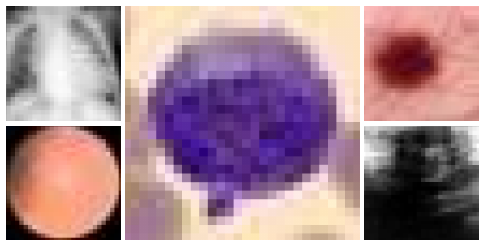}
    \caption{}
    \label{fig:reconstructions_orig}
\end{subfigure}
\hfill
\begin{subfigure}{0.32\textwidth}
    \includegraphics[width=\textwidth]{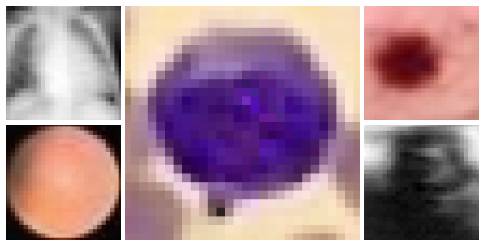}
    \caption{}
    \label{fig:reconstructions_AE}
\end{subfigure}
\caption{\textbf{a)} Reconstruction unORANIC \textbf{b)} Original input \textbf{c)} Reconstructions AE.}
\label{fig:reconstructions}
\end{figure}

We compared unORANIC's reconstruction results with a vanilla autoencoder (AE) architecture to assess its encoding and reconstruction abilities. The vanilla AE shares the same architecture and latent dimension as the anatomy branch of our model. The average peak signal-to-noise ratio (PSNR) values for the reconstructions of both methods on the selected datasets are presented in \tablename~\ref{tab:reconstructions}. Both models demonstrate precise reconstruction of the input images, with our model achieving a slight improvement across all datasets. This improvement is attributable to the concatenation of anatomy and characteristic features, resulting in twice the number of features being fed to the decoder compared to the AE model. The reconstruction quality of both models is additionally depicted visually in \figurename~\ref{fig:reconstructions} using selected examples.

\begin{table}
\caption{PSNR values of the reconstructions}
\label{tab:reconstructions}
    \centering
    \begin{tabular}{l c c c c c c c c c c }
        \toprule
        Methods & \phantom{xxxx} & Blood & \phantom{xxxx} & Breast & \phantom{xxxx} & Derma & \phantom{xxxx} & Pneumonia & \phantom{xxxx} & Retina \\
        \midrule
        AE & & $30.837$ & & $28.725$ & & $38.125$ & & $34.787$ & & $35.872$ \\
        unORANIC & & $\mathbf{31.700}$ & & $\mathbf{29.390}$ & & $\mathbf{38.569}$ & & $\mathbf{36.040}$ & & $\mathbf{36.309}$\\
        \bottomrule
    \end{tabular}
\end{table}

\subsection{Corruption Revision}
\label{subsec:corruption_revision}
\begin{figure}
\centering
\begin{subfigure}{0.55\textwidth}
    \includegraphics[width=\textwidth]{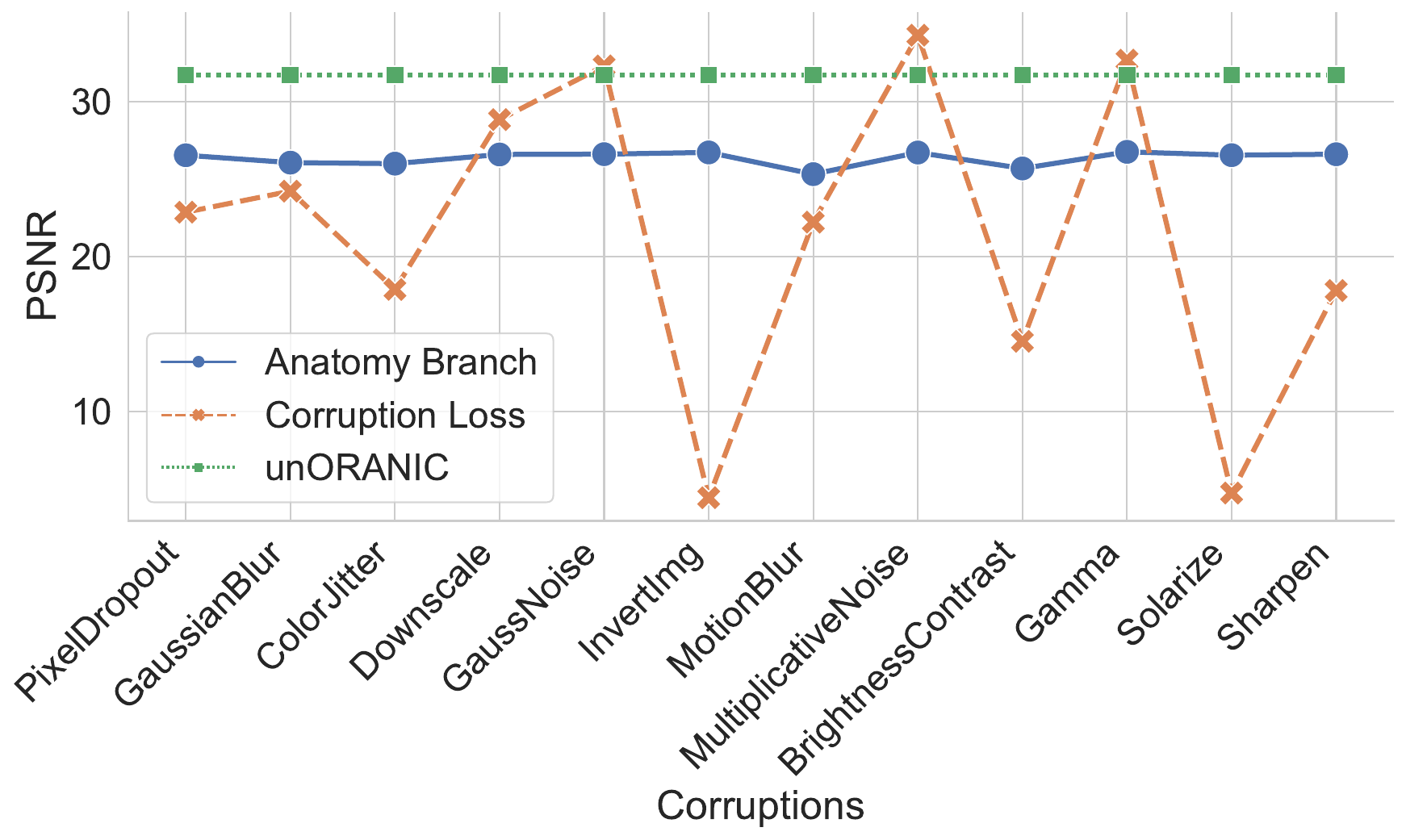}
    \caption{}
    \label{fig:reconstructions_corruptions_plot}
\end{subfigure}
\hfill
\begin{subfigure}{0.43\textwidth}
    \includegraphics[width=\textwidth]{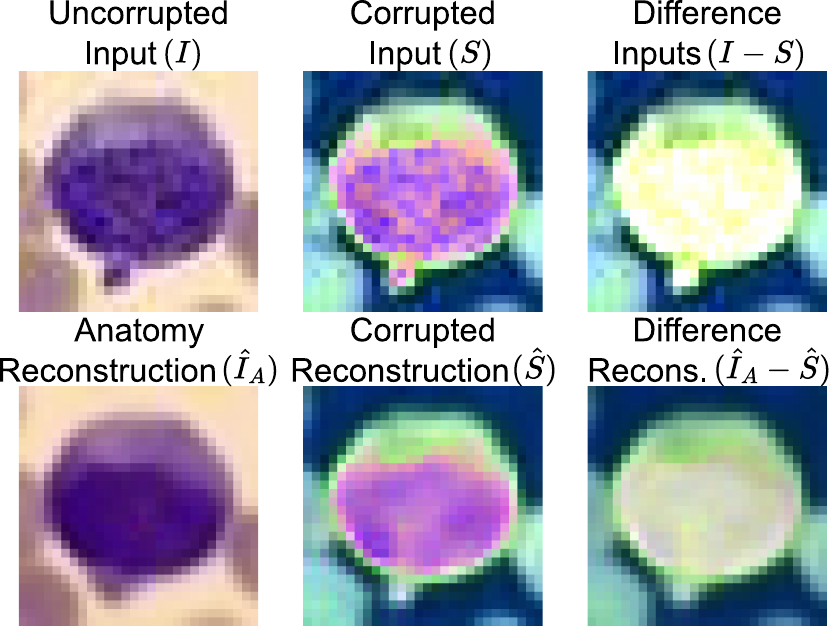}
    \caption{}
    \label{fig:reconstructions_corruptions}
\end{subfigure}
\caption{\textbf{a)} Corruption revision results of unORANIC for a set of corruptions. \\ \textbf{b)} Visualization of unORANIC's corruption revision for a solarization corruption. The method reconstructs $I$ via $\hat{I}_A$ while seeing only the corrupted input $S$.}
\label{fig:reconstruction_with_corruptions}
\end{figure}

We now evaluate the capability of unORANIC's anatomical branch to revise existing corruptions in an input image. For this, all input images in the test set $\left\{I_1, I_2, ..., I_n\right\}$ are intentionally corrupted using a set of corruptions from the Albumentations library \cite{Buslaev2020} to generate synthetic distorted versions $\left\{S_1, S_2, ..., S_n\right\}$ first, before passing those through the unORANIC network afterward. Each corruption is hereby uniformly applied to all test images. Despite those distortions, the anatomy branch of the model should still be able to reconstruct the original, uncorrupted input images $\hat{I}_i$. To assess this, we compute the average PSNR value between the original images $I_i$ and their corrupted versions $S_i$. Afterward, we compare this value to the average PSNR between the original uncorrupted input images $I_i$ and unORANIC's anatomical reconstructions $\hat{I}_{A_i}$. Both of these values are plotted against each other in \figurename~\ref{fig:reconstructions_corruptions_plot}, for the BloodMNIST dataset. For reference purposes, the figure contains the PSNR value for uncorrupted input images as well. It can be seen that the anatomical reconstruction quality is close to the uncorrupted one and overall consistent across all applied corruptions which proves unORANIC's corruption revision capability. As it can be seen in \figurename~\ref{fig:reconstructions_corruptions}, this holds true even for severe corruptions such as solarization.

\subsection{Classification}
We proceed to assess the representativeness of the encoded feature embeddings, comprising both anatomical and characteristic information, and their suitability for downstream tasks or applications. To this end, we compare our model with the ResNet-18 baseline provided by MedMNIST v2 on two distinct tasks. The first task involves the classification of each dataset, determining the type of disease (e.g., cancer or non-cancer). For the second task, the models are assigned to detect whether a corruption (the same ones used in the previous revision experiment) has or has not been applied to the input image in the form of a binary classification. To do so, we freeze our trained network architecture and train a single linear layer on top of each embedding for its respective task. The same procedure is applied to the AE architecture, while the MedMNIST baseline architecture is retrained specifically for the corruption classification task. Results are summarized in \tablename~\ref{tab:classification_task} and \tablename~\ref{tab:detection_corruption}, with AUC denoting the Area under the ROC Curve and ACC representing the Accuracy of the predictions. Overall, our model's classification ability is not as strong as that of the baseline model. However, it is important to note that our model was trained entirely in an unsupervised manner, except for the additional linear layer, compared to the supervised baseline approach. Regarding the detection of corruptions within the input images, our model outperforms the baseline method. Furthermore, it should be emphasized that our model was trained simultaneously for both tasks, while the reference models were trained separately for each individual task.

\begin{table}
\caption{Classification results on each selected dataset}
\label{tab:classification_task}
    \centering
    \begin{tabular}{l c c c c c c c c c c c c c c c}
        \toprule
        \multirow{2.5}{*}{Methods} & \phantom{i} & \multicolumn{2}{c}{Blood} & \phantom{i} & \multicolumn{2}{c}{Breast} & \phantom{i} & \multicolumn{2}{c}{Derma} & \phantom{i} & \multicolumn{2}{c}{Pneumonia} & \phantom{i} & \multicolumn{2}{c}{Retina} \\
        \cmidrule{3-4} \cmidrule{6-7} \cmidrule{9-10} \cmidrule{12-13} \cmidrule{15-16}
        & & AUC & ACC & & AUC & ACC & & AUC & ACC & & AUC & ACC & & AUC & ACC \\
        \midrule
        Baseline & & $\mathbf{0.998}$ & $\mathbf{0.958}$ & & $\mathbf{0.901}$ & $\mathbf{0.863}$ & & $\mathbf{0.917}$ & $\mathbf{0.735}$ & & $0.944$ & $0.854$ & & $\mathbf{0.717}$ & $0.524$ \\
        AE & & $0.966$ & $0.806$ & & $0.819$ & $0.801$ & & $0.787$ & $0.694$ & & $0.904$ & $0.840$ & & $0.678$ & $0.470$ \\
        unORANIC & & $0.977$ & $0.848$ & & $0.812$ & $0.808$ & & $0.776$ & $0.699$ & & $\mathbf{0.961}$ & $\mathbf{0.862}$ & & $0.691$ & $\mathbf{0.530}$\\
        \bottomrule
    \end{tabular}
\end{table}

\begin{table}
\caption{Corruption detection on each selected dataset}
\label{tab:detection_corruption}
    \centering
    \begin{tabular}{l c c c c c c c c c c c c c c c}
        \toprule
        \multirow{2.5}{*}{Methods} & \phantom{i} & \multicolumn{2}{c}{Blood} & \phantom{i} & \multicolumn{2}{c}{Breast} & \phantom{i} & \multicolumn{2}{c}{Derma} & \phantom{i} & \multicolumn{2}{c}{Pneumonia} & \phantom{i} & \multicolumn{2}{c}{Retina} \\
        \cmidrule{3-4} \cmidrule{6-7} \cmidrule{9-10} \cmidrule{12-13} \cmidrule{15-16}
        & & AUC & ACC & & AUC & ACC & & AUC & ACC & & AUC & ACC & & AUC & ACC \\
        \midrule
        Baseline & & $0.548$ & $0.184$ & & $0.565$ & $0.571$ & & $0.517$ & $0.046$ & & $0.614$ & $0.524$ & & $0.500$ & $0.283$ \\
        AE & & $0.746$ & $0.736$ & & $0.576$ & $0.545$ & & $\mathbf{0.655}$ & $\mathbf{0.640}$ & & $0.648$ & $0.657$ & & $0.827$ & $0.785$ \\
        unORANIC & & $\mathbf{0.755}$ & $\mathbf{0.746}$ & & $\mathbf{0.612}$ & $\mathbf{0.590}$ & & $0.643$ & $0.620$ & & $\mathbf{0.667}$ & $\mathbf{0.660}$ & & $\mathbf{0.847}$ & $\mathbf{0.823}$\\
        \bottomrule
    \end{tabular}
\end{table}

\subsection{Corruption Robustness}
\begin{figure}[t]
\centering
\begin{subfigure}{0.42\textwidth}
    \includegraphics[width=\textwidth]{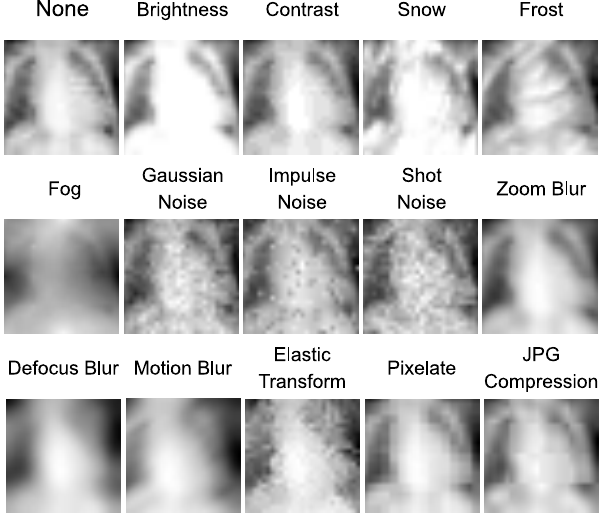}
    \caption{}
    \label{fig:imagecorruptions}
\end{subfigure}
\hfill
\begin{subfigure}{0.55\textwidth}
    \includegraphics[width=\textwidth]{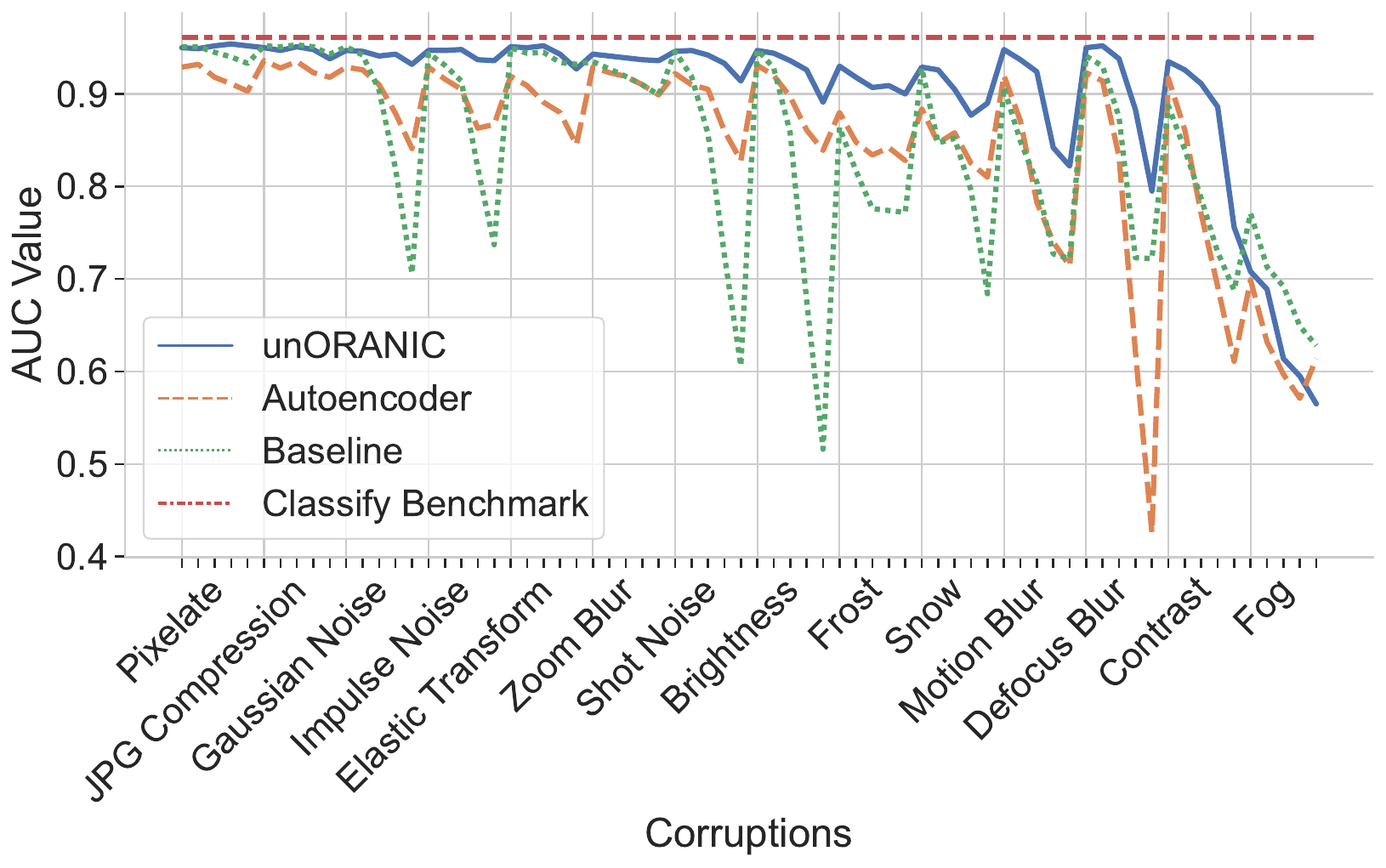}
    \caption{}
    \label{fig:corruption_robustness_plot}
\end{subfigure}
\caption{\textbf{a)} Visualization of the applied corruptions. \textbf{b)} Classification results of unORANIC compared to the vanilla AE and the baseline when exposed to the applied corruptions with increasing severity.}
\label{fig:corruption_robustness}
\end{figure}

Lastly, we assess unORANIC's robustness against unseen corruptions during training. To accomplish this, we adopt the image corruptions introduced by \cite{hendrycks2019robustness,michaelis2019dragon}, following a similar methodology as outlined in Section~\ref{subsec:corruption_revision}. The used corruptions are presented in \figurename~\ref{fig:imagecorruptions} for an example of the PneumoniaMNIST dataset. Sequentially, we apply each corruption to all test images first and subsequently pass these corrupted images through the trained unORANIC, vanilla AE and baseline model provided by the MedMNIST v2 benchmark for the datasets associated classification task. This process is repeated for each individual corruption. Additionally, we vary the severity of each corruption on a scale of 1 to 5, with 1 representing minor corruption and 5 indicating severe corruption. By collecting the AUC values for each combination of corruption and severity on the PneumoniaMNIST dataset and plotting them for each method individually, we obtain the results presented in \figurename~\ref{fig:corruption_robustness_plot}. Notably, our model demonstrates greater overall robustness to corruption, particularly for noise, compared to both the baseline and the simple autoencoder architecture.

\section{Conclusion}
The objective of this study was to explore the feasibility of orthogonalizing anatomy and image-characteristic features in an unsupervised manner, without requiring specific dataset configurations or prior knowledge about the data. The conducted experiments affirm that unORANIC's simplistic network architecture can explicitly separate these two feature categories. This capability enables the training of corruption-robust architectures and the reconstruction of unbiased anatomical images. The findings from our experiments motivate further investigations into extending this approach to more advanced tasks, architectures as well as additional datasets, and explore its potential for practical applications.


%
%
%
\bibliographystyle{splncs04}
\bibliography{mybibliography}
%




\end{document}